# Deterministic Domain Wall Motion Orthogonal To Current Flow Due To Spin Orbit Torque

Debanjan Bhowmik[1], Mark E. Nowakowski[1], Long You[1], OukJae Lee[1], David Keating[2], Mark Wong[1], Jeffrey Bokor[1,3] and Sayeef Salahuddin[1,3,*]

[1]Department of Electrical Engineering and Computer Sciences, University of California Berkeley, Berkeley, CA 94709, USA

[2]Department of Physics, University of California Berkeley, Berkeley, CA 94709, USA

[3]Material Science Division, Lawrence Berkeley National Laboratory

Abstract- Deterministic control of domain walls orthogonal to the direction of current flow is demonstrated by exploiting spin orbit torque in a perpendicularly polarized Ta/CoFeB/MgO multilayer in presence of an in-plane magnetic field. Notably, such orthogonal motion with respect to current flow is not possible from traditional spin transfer torque driven domain wall propagation even in presence of an external magnetic field. Reversing the polarity of either the current flow or the in-plane field is found to reverse the direction of the domain wall motion. From these measurements, which are unaffected by any conventional spin transfer torque by symmetry, we estimate the spin orbit torque efficiency of Ta to be 0.08.

* All correspondence to be addressed to Sayeef Salahuddin (email: sayeef@berkeley.edu)

According to the conventional spin transfer torque model, when spin polarized electrons coming from a magnetized region impinge on the spins in a domain wall, they exert a torque that tries to orient the domain wall spins in the direction of the incoming spins [1-4]. The strength of the torque is proportional to the relative angle between the incoming spins and domain wall spins. If a domain wall is formed along the width of a perpendicularly polarized magnetic wire (transverse domain wall) as shown in Fig. 1(a), then a current flowing in the long direction can exert such a torque and move the domain wall. This has been the canonical configuration for studying current induced domain wall motion [1]. If, on the other hand, a domain wall is formed along the length as shown in Fig. 1(b) and Fig. 1(d), then no conventional spin transfer torque, to be referred to as the bulk spin torque, is exerted on this longitudinal wall by a current that flows along the length, even in the presence of an external in-plane magnetic field [5]. This is because the domain wall spins do not change direction along the path of the current in this case. The situation is different in a perpendicularly polarized magnetic heterostructure where an underlying non-magnetic heavy metal can provide a spin orbit torque [6-9,11-20]. Here the torque originates from the accumulation of spin-polarized electrons at the interface whose direction of polarization is different from the magnet and is determined by the direction of current flow and the sign of the spin-orbit coefficient. For example, in a heterostructure of Ta (heavy metal) /CoFeB (ferromagnet), a current flowing in the longitudinal +x direction would accumulate –y polarized spins at the Ta/CoFeB interface (Fig. 1(c)). The magnet (CoFeB) itself is polarized in the ±z direction. As a result, there is always a relative angle between magnetic polarization and accumulated spins and hence a torque is exerted on the magnet independent of the specific topology of the magnetic domain wall (longitudinal or transverse). Indeed, here we show that a longitudinal current can deterministically move a longitudinal domain wall orthogonal to the current flow in the presence of an in-plane magnetic field. Reversing the direction of the current flow or the direction of the field reverses the direction of the domain wall motion.

For experimental investigation, Hall bars were fabricated from a stack of Si (substrate)/ SiO$_2$ (100 nm)/ Ta (10 nm)/ CoFeB (1 nm)/ Ta (2 nm), exhibiting perpendicular magnetic anisotropy [5,10]. Current is applied along the bar of width 20 microns, which is along the x-axis, and anomalous Hall voltage is measured in the y direction across the narrower bar of width 5 microns [Fig. 1(d)] to obtain the anomalous Hall resistance ($R_{AHE}$). The anomalous Hall resistance ($R_{AHE}$) versus magnetic field plot of Fig. 2(a) shows that the CoFeB layer exhibits perpendicular anisotropy with a magnetic field of ~30 G needed to switch its magnetization between the two saturated states- "into the plane" or $m_z=1$ state ($\otimes$) with $R_{AHE}=1.2\ \Omega$ and "out of the plane" or $m_z=-1$ state ($\odot$) with $R_{AHE}=-1.2\ \Omega$; $m_z$ is the magnetization in z-direction, normalized by the saturation magnetization. Along with measurement of $R_{AHE}$, Magneto Optic Kerr Effect (MOKE) has been used to image the magnetic domains of the wider Hall bar [6,7,8]. To obtain the best contrast, each image has been subtracted from a reference image of a saturated "into the plane" state of the magnet. The subtracted image of the saturated "into the plane" state of the magnet shows no contrast between the magnetic bar and the background substrate material because the substrate does not contribute to a magnetic signal and the positive magnetic signal of the "into the plane" magnet when subtracted from another image of "into the plane" magnet yields zero (Fig. 2(a)). On the other hand, when an image of a saturated "out of the plane" state is subtracted from the reference image of "into the plane" state, the substrate still gives no signal, but the difference between negative signal of the "out of the plane" magnet and positive signal of the "into the plane" magnet is non-zero. So a subtracted image of "out of the plane" state shows a dark contrast between the magnetic bar and the background (Fig. 2(b)).

Starting from the magnet saturated "into the plane" ($m_z=1$, $R_{AHE}=1.2\ \Omega$), a current pulse of magnitude $7.5\times10^6$ A/cm$^2$ is applied along the wider Hall bar in +x direction at a zero magnetic field [Fig. 1(d)]. At the end of the pulse, the steady state $R_{AHE}$ of the final magnetic state is measured to be ~0 $\Omega$ [Fig. 2(a)]. The MOKE image of that state shows the magnet to be split into a domain of "out of the plane" or $m_z=-1$ ($\odot$) polarized moments for y<0 and a domain of "into the plane" or $m_z=1$ ($\otimes$) polarized moments for y>0 with a longitudinal domain wall separating the two, based on the chosen coordinate system [Fig. 2(a) and Fig. 1(d)]. We call this state the "mixed" state [Fig. 2(a)] because the magnet has both $m_z=1$ and $m_z=-1$ polarized domains in this state. Reversing the polarity of the current reverses the position of the two domains in the "mixed" state [Fig. 2(c)]. The polarity of the domains in the "mixed" state follows the out of the plane component of the Oersted field, generated by the current pulse, at the edges of the bar. Micromagnetic simulations [5] show that starting from a saturated state, reverse polarized domains are nucleated at the edge of the bar due to the Oersted field. Subsequently, the domain wall moves from the edge of the bar to the center to reduce the magnetostatic energy of the system. Once the longitudinal domain wall is formed at the center of the bar after the application of the current pulse, application of an out of plane magnetic field moves the domain wall in one direction or another based on the polarity of the field as observed from the MOKE images [Fig. 2(b)]. The displacement of the domain wall is proportional to the magnitude of the applied field, indicating that the domain wall motion is governed by pinning defects [21, 22]. We also observe that the domain wall starts moving from the center of the bar only when the out of plane field is 15 G or above. If the magnetic field is next applied at different angles (θ) with respect to the film normal, the magnetic field needed to move the domain wall follows a 1/cos(θ) dependence as expected from the Knodorsky model of domain wall depinning field [21].

Creation of such a longitudinal domain wall provides us with the unique opportunity to control its motion orthogonal to the current flow using spin orbit torque, which is not otherwise possible from a bulk spin torque. Deterministic control of the motion of this domain wall orthogonal to the current flow is the central point of the paper, which we discuss next. We first saturate the magnetic bar in "into the plane" ($m_z$=1) state and apply a current pulse of magnitude $7.5 \times 10^6$ A/cm$^2$ along the bar in +x direction at zero magnetic field to create a "mixed" state with a longitudinal domain wall at the center of the bar just as in Fig. 2(a). Starting from the longitudinal domain wall, a current pulse is applied along the bar in the +x direction in the presence of an in-plane magnetic field opposite to it [Fig. 3(a)]. This is repeated for different magnitudes of the current pulse. MOKE images of the bar taken after every current pulse show that the current pulse moves the domain wall in -y direction such that the "into the plane" or $m_z$=1 polarized domain expands while the "out of the plane" or $m_z$=-1 polarized domain contracts. The distance moved by the domain wall is proportional to the magnitude of the current pulse. Reversing the direction of the magnetic field reverses the direction of the domain wall motion. We next vary the in-plane magnetic field and apply current pulses of different magnitude starting every time from the same initial condition of a longitudinal domain wall at the center of the bar just as in "mixed state" of Fig. 2(a). We measure $R_{AHE}$, indicative of the domain wall position, after each current pulse and plot the $R_{AHE}$ as a function of the current and the in-plane field. The contour plot obtained in that process [Fig. 3(b)] shows that when the in-plane magnetic field is positive (along +x) current along +x (positive polarity) beyond a threshold value (~5 × $10^6$ A/cm$^2$) moves the domain wall in +y direction and hence the average $m_z$ <0 ($R_{AHE}$ < 0, blue color in contour plot). The higher the current or stronger the field, the more negative is the average $m_z$ and the $R_{AHE}$. In the presence of a positive in-plane magnetic field (along +x), current in –x direction (negative polarity) moves the domain wall in -y direction and the average $m_z$>0 ($R_{AHE}$>0, red color in contour plot). When in-plane magnetic field is applied in –x direction (negative field), positive current pulse moves the domain wall in -y direction ($R_{AHE}$>0, red color) and negative current pulse moves the domain wall in +y direction ($R_{AHE}$<0, blue color).

We performed micromagnetic simulations in Object Oriented Micromagnetic Framework to explain this result [23,24]. Starting from a longitudinal domain wall at the center of a rectangular magnet of length 600 nm, width 200 nm and thickness 1 nm, the system is allowed to evolve with time under the influence of an in-plane magnetic field along its length and a Slonczewski-like spin orbit torque [25] orthogonal to it. The direction of the applied in-plane magnetic field in the experiment (Fig. 3 and Fig. 4) and the shape anisotropy owing to the length of the longitudinal domain wall favor the formation of a Bloch wall. Dzyaloshinskii-Moriiya Interaction (DMI) [7] is found to be small in our samples (see supplemenetary material [5]). In fact, the applied magnetic field orients the moments in the domain wall in its direction and provides directionality to the domain wall motion. When a magnetic field is applied in -x direction and a current is applied in the +x direction the domain wall moves in -y direction till the magnet is saturated into the plane [Fig. 4(a)]. Reversing the direction of the magnetic field [Fig. 4(b)] or the current [Fig. 4(c)] reverses the direction of domain wall motion, just as observed experimentally. Thus, polarity of the final state of the magnet is found to be a cross product of direction of the in-plane field and direction of accumulated spins. Notably, these simulations results which show a one-to-one correspondence with the experimental observation are consistent with the Slonczewski-like spin orbit torque.

Oersted field or field like spin orbit torque from the current cannot explain such domain wall motion [5,13,25].

One way to make a quantitative estimate for the efficiency of the spin-orbit torque is to relate the current density needed to move the domain walls to the effective out-of plane field, following Thiaville *et al.* [26]. Starting from the same initial condition of a longitudinal wall at the center of the magnetic bar identical to the "mixed state" of Fig. 2(a), current pulses of different magnitude are applied along the bar in the +x direction with a magnetic field applied in -x direction (Fig. 3). $R_{AHE}$ is measured after each pulse to indicate the position of the domain wall. The same experiment is repeated for different values of in-plane magnetic field. We observe that for very small in-plane fields current pulses up to the magnitude of $8\times10^6$ A/cm$^2$ move the domain wall barely [Fig. 3(c)]. For an in-plane magnetic field of magnitude 45 G and above, the distance moved by the domain wall, is linearly proportional to the current density ($J_c$) of the applied pulse. As a result the $R_{AHE}$ measured after the current pulse varies linearly with current density till the domain wall moves to the edge of the bar to switch the entire magnet and hence $R_{AHE}$ reaches saturation. The slope of the linear curve ($\frac{\partial R_{AHE}}{\partial J_C}$) is approximately the same for different magnitudes of in-plane magnetic field above 45 G and is equal to $7.39\times10^{-7}$ Ω/(A/cm$^2$) [5]. Fig. 2(a) shows that the longitudinal domain wall moves under the application of an out of plane field such that the $R_{AHE}$ varies linearly with the out of plane field at a rate of 0.15 Ω/G ($\frac{\partial R_{AHE}}{\partial H_{out}}$). Comparing the two slopes, we conclude that for an in-plane field sufficient enough to switch the net magnetic moment in the domain wall in the direction of the field so that the moment is orthogonal to the direction of spin polarization of the accumulated electrons [7,26,27,28] the current ($J_c$) acts as an effective out of plane field ($H_{out}$) such that $\frac{\partial H_{out}}{\partial J_C} = (\frac{\partial R_{AHE}}{\partial J_C})/(\frac{\partial R_{AHE}}{\partial H_{out}}) = 4.92\times10^{-6}$ G/(A/cm$^2$). We can relate this slope to the expression $H_{out} = \frac{\pi}{2}\frac{\hbar\theta}{2et_FM_s}$, [26] where θ is the efficiency of spin orbit torque, e is the charge of an electron, $\hbar$ is the Planck constant, $M_s$ is the saturated magnetization of the ferromagnet (800 e.m.u/c.c.), as measured for our thin film stack by Vibrating Sample Magnetometry) and $t_F$ is the thickness of the ferromagnet (1 nm). This provides θ = 0.076, which is comparable to the spin Hall angle (0.12) reported by L. Liu *et al.* [12] in a similar heterostructure.

To summarize, we have shown that spin orbit torque can be exploited to deterministically move a domain wall in a perpendicularly polarized magnet orthogonal to current flow, which is otherwise not achievable by conventional bulk spin torque. Reversing the polarity of the current or the in-plane magnetic field reverses the direction of motion of the domain wall. This adds a new capability to the toolset of current induced domain wall control and can impact the way a domain wall is routed through complicated structures [29,30]. The specific configuration presented in this work also provides a way to investigate the spin-orbit torque directly, unaffected by any bulk spin torque which is zero by symmetry.

We acknowledge Giovanni Finocchio, Satoru Emori, Plamen Stamenov, Michael Coey, Dominic Labanowski and Mihika Prabhu for valuable discussions. This work was supported in part by NSF Grants 1111733, 0939514, 1017575 and by the STARNET FAME Center.

**Figure Captions**

FIG. 1 **(**a) Current flows across a transverse domain wall. Green arrows represent the magnetic moments. (b) Current flows along a longitudinal domain wall. (c) When current flows through the Ta in x direction (electrons in –x direction) electrons with spin polarization in -y direction accumulate at the interface of the Ta and CoFeB. This results in the transfer of spin orbit torque to the domain wall in CoFeB. **(**d) The MOKE image of the 20 microns wide magnetic bar shows the formation of a longitudinal wall.

FIG. 2 **(**a) Anomalous Hall resistance ($R_{AHE}$) versus magnetic field plot along with MOKE images of a saturated "into the plane" ($\otimes$) polarized state ($m_z$=1 or $R_{AHE}$=1.2 $\Omega$) and a saturated "out of the plane" ($\odot$) polarized state ($m_z$=-1 or $R_{AHE}$=-1.2 $\Omega$) polarized state. Starting from the $m_z$=1 state application of a current pulse of magnitude $7.5\times10^6$ A/cm$^2$ at a zero magnetic field causes the formation of the "mixed state" ($R_{AHE} \sim 0$ $\Omega$). MOKE image of the "mixed state" shows a longitudinal domain wall. Under the application of a magnetic field in +z direction, $R_{AHE}$ increases till the magnet reaches the saturated "into the plane" state (red plot). Applying a magnetic field in -z direction causes reduction in $R_{AHE}$ till the magnet gets saturated in the "out of plane" direction (blue plot). (b) MOKE images show the movement of the longitudinal domain wall from the center to the edge of the bar under the application of +z and –z magnetic fields. (c) Mixed state, formed by a positive current pulse (current I along +x) has "out of the plane" ($\odot$)

polarized domains for y<0 and "into the plane" (⊗) polarized domains for y>0 while mixed state, formed by a negative current pulse (I along –x) has "into the plane" (⊗) polarized domains for y<0 and "out of the plane" (⊙) polarized domains for y>0. (d) Minimum field needed to move the longitudinal domain wall from the center of the bar is plotted against the angle of application of the field.

FIG. 3 (a) MOKE images of the magnetic bar after a positive current pulse of different magnitude is applied on a longitudinal domain wall ["mixed" state of Fig. 2(a)] in the presence of an in-plane field. (b) Contour plots of final $R_{AHE}$ after a current pulse versus the magnitude of the current and magnitude of the applied in-plane field for different combinations of positive and negative currents and fields. (c) Plots of $R_{AHE}$ versus magnitude of current pulse for different values of magnetic field, applied in –x direction.

FIG. 4 (a), (b) Micromagnetic simulation of a 600 nm long and 200 nm wide magnet shows the motion of a longitudinal domain wall under the application of a spin polarization $\vec{\sigma}$ in –y direction (current $I$ in +x direction) and the magnetic field H in –x/+x direction. The direction of the net magnetic moment in the domain wall, shown by a black arrow, is initialized along the direction of applied field. (c) Micromagnetic simulation of the same magnet with spin polarization in +y direction (current I in –x direction) and magnetic field in –x direction.

# Figures

Figure 1

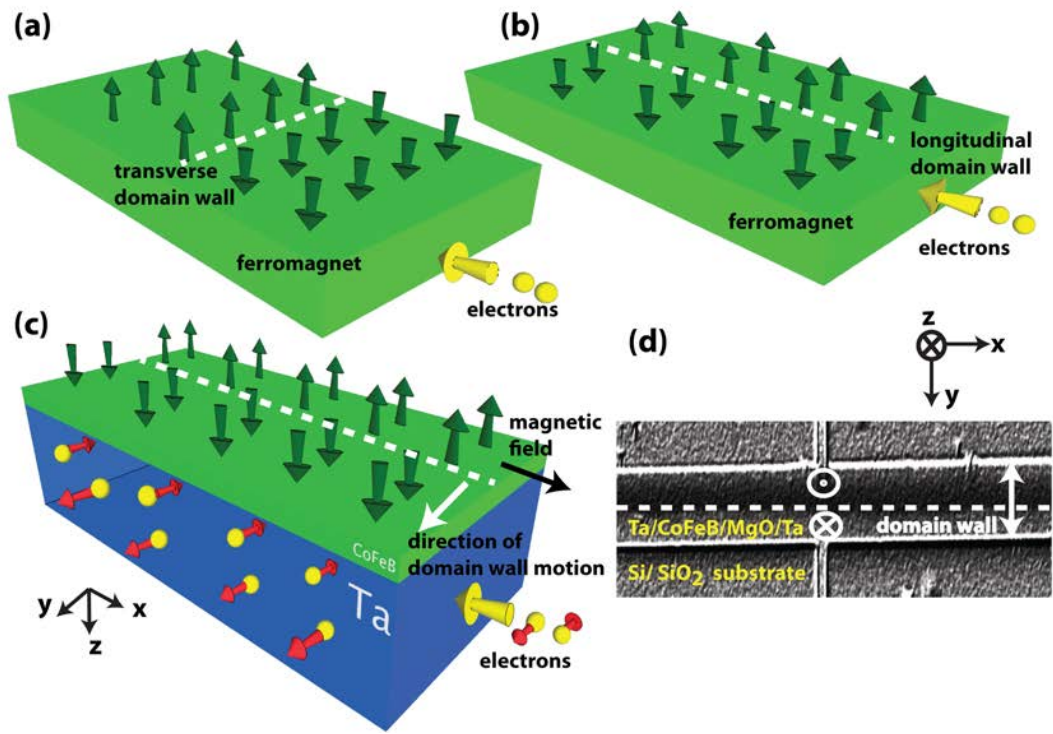

Figure 2

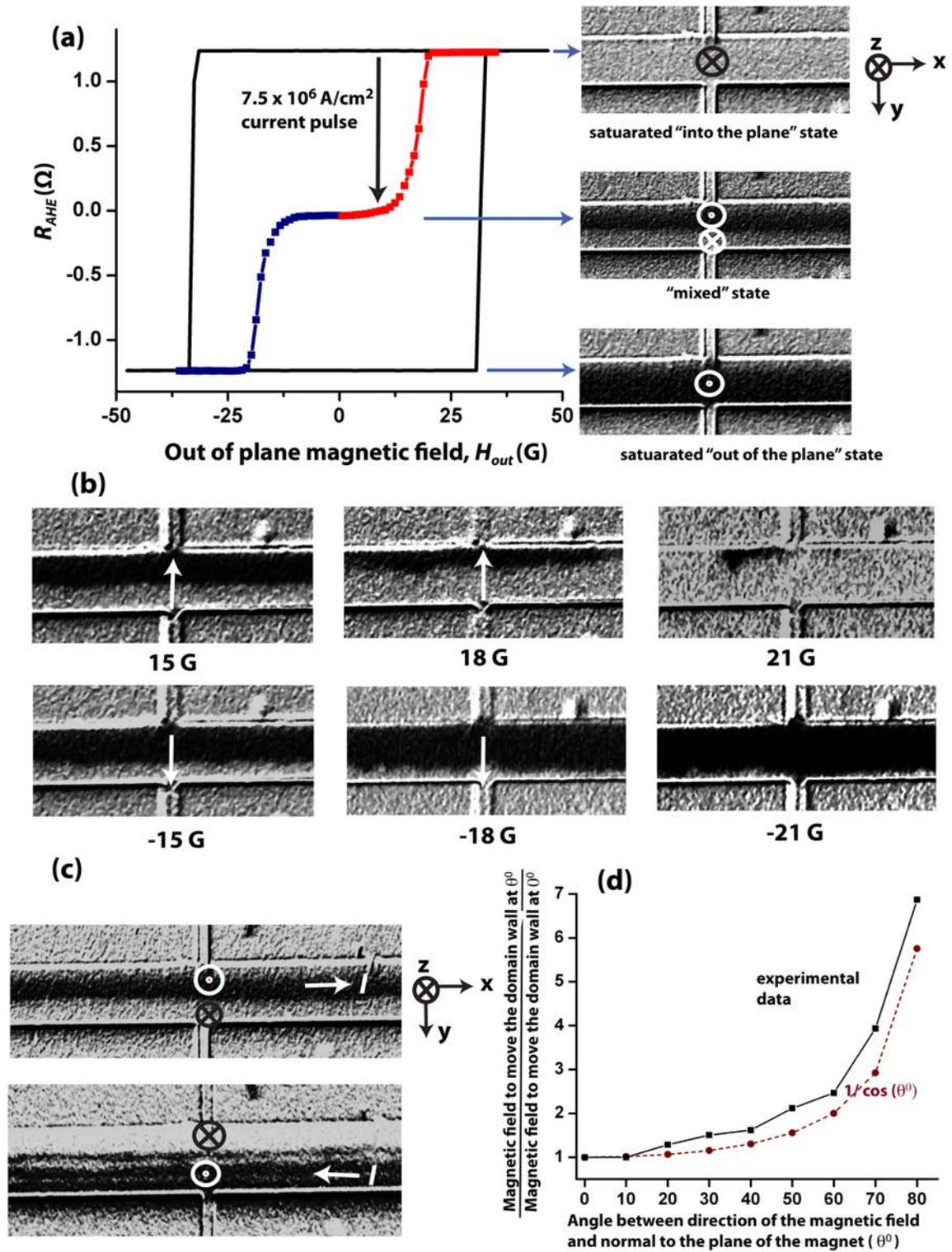

Figure 3

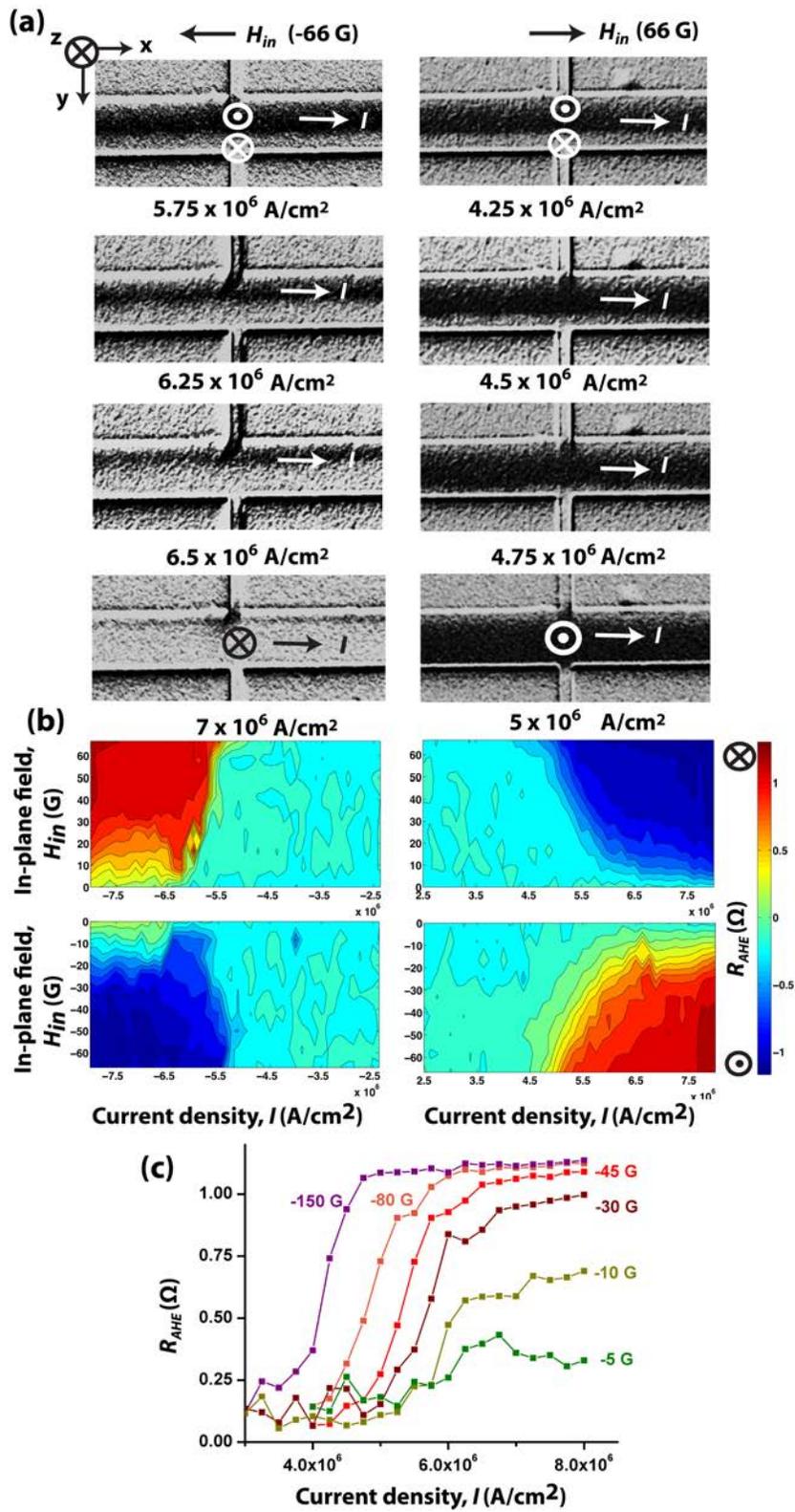

Figure 4

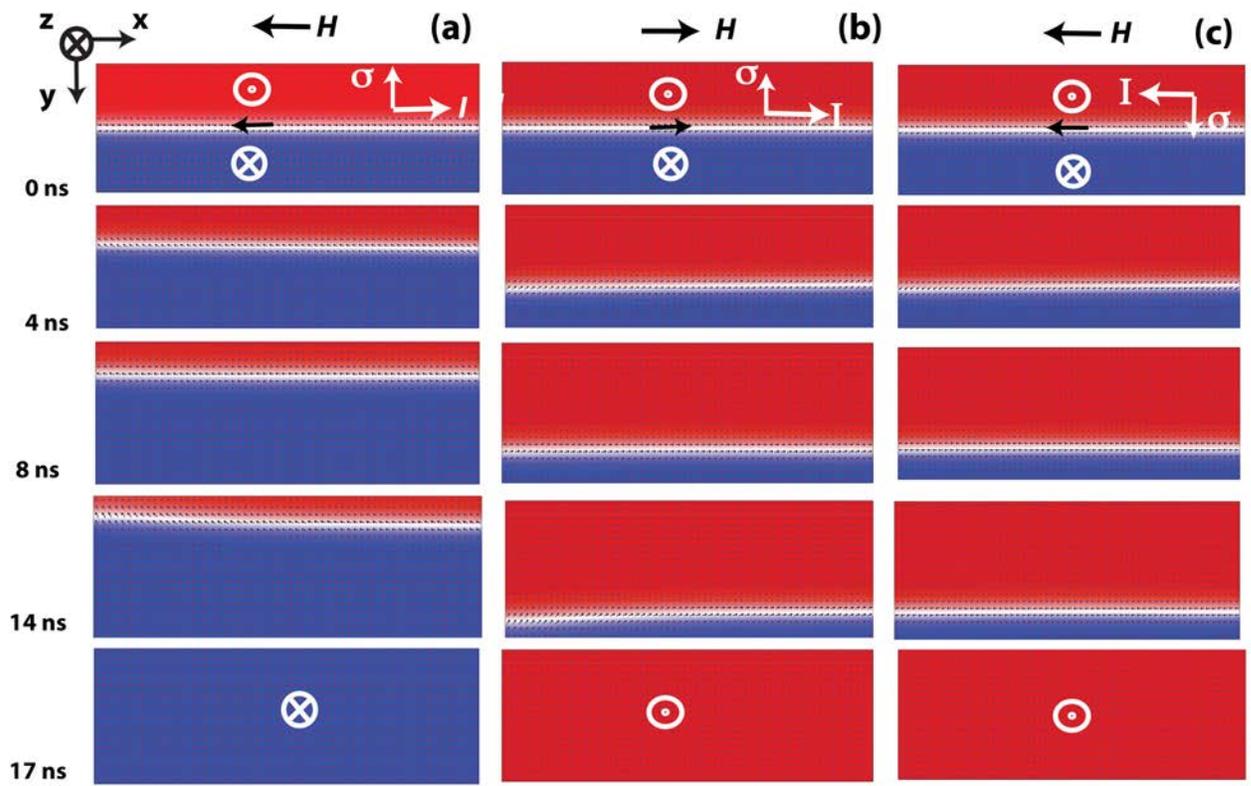

# "Deterministic Domain Wall Motion Orthogonal To Current Flow Due To Spin Orbit Torque"

Debanjan Bhowmik[1], Mark E. Nowakowski[1], Long You[1], OukJae Lee[1], David Keating[2], Mark Wong[1], Jeffrey Bokor[1,3] and Sayeef Salahuddin[1,3]

[1]Department of Electrical Engineering and Computer Sciences, University of California Berkeley, Berkeley, CA 94709, USA

[2]Department of Physics, University of California Berkeley, Berkeley, CA 94709, USA

[3]Material Science Division, Lawrence Berkeley National Laboratory

**Section S1- Micromagnetic simulation showing that conventional/ bulk spin torque can move the transverse domain wall but not the longitudinal domain wall**

When current flows through a ferromagnet, conduction electrons in the ferromagnet apply a "bulk spin torque" on the domain wall of the ferromagnet, given by $b_J(\vec{J}_C.\nabla)\vec{m} - \beta b_J \vec{m} \times ((\vec{J}_C.\nabla)\vec{m})$, where $b_J = \frac{\mu_B P}{e M_s} J_c$; $\mu_B$- Bohr magnetron, P- spin polarization, e- charge of an electron, $M_s$- saturation magnetization of the ferromagnet, β- non-adiabatic parameter and $J_c$- charge current [1-4].

For a transverse domain wall, since the current flows across the domain wall, the magnetization has a non-zero gradient along the direction of the charge current ($(\vec{J}_C.\nabla)\vec{m} \neq 0$). So the bulk spin torque is non-zero. We perform micromagnetic simulations on Object Oriented MicroMagnetic Framework (OOMMF) [5] using the extension module for current induced domain wall motion (class spinTEevolve) [6] to show that a transverse domain wall moves due to the bulk spin torque when current flows across it. Only bulk spin torque and no spin orbit torque is present in this simulation (Fig. S1(a)).

A 600 nm long, 200 nm wide and 1 nm thick magnet is used for simulations, with a mesh size of 2 nm laterally and 1 nm across the thickness. For all micromagnetic simulation figures in the main paper and the supplementary material, blue dots with red background represent moments pointing out of the plane (-z) while red dots with blue background represent moments pointing into the plane (+z). The simulation parameters are: saturation magnetization $M_s$= 8 x $10^5$ A/m (measured for our materials stack by performing Vibrating Sample Magnetometry), exchange constant A= 3 x $10^{-11}$ J/m, perpendicular anisotropy constant K= 6 x $10^5$ J/m$^3$, spin polarization P=0.5, non-adiabatic parameter β=0.04 and charge current $J_c$= $10^8$ A/cm$^2$. Starting from a transverse domain wall at the center of the magnet, we see that the domain wall moves with time under the application of current due to the bulk spin torque (Fig. S1(a)). The domain wall moves in the direction of propagation of electrons, i.e., opposite to the direction of the current, as expected from theory [1-3].

When we simulate a longitudinal domain wall instead, current flows along the domain wall. In this case the direction, in which the gradient of the magnetization is not zero, and direction of the current are orthogonal, so $(\vec{J_c} \cdot \nabla)\vec{m} = 0$. So the longitudinal domain wall does not experience a bulk spin torque and hence it does not move under the application of a current pulse, as we verify through micromagnetic simulations using the same simulation parameters as the transverse domain wall case (Figure S1(b)). Only bulk spin torque and no spin orbit torque is present in this simulation unlike the simulations of Figure 4 of the main paper.

In our experiments we move the longitudinal domain wall with current pulses in the presence of an in-plane magnetic field along the current direction (Figure 3 and Figure 4 of the main text). In Figure S1c, we show simulations of a bulk spin torque acting on a longitudinal domain wall just like Fig. S1(b), with an in-plane magnetic field of 10G applied in addition to it along the current direction. We see that even in this case the longitudinal domain wall does not move with time. Fig. S1(d) shows that when the magnetic field is orthogonal to the current, the bulk spin torque does not move the domain wall either. Again, no spin orbit torque is considered in these simulations. Thus we conclude that the bulk spin torque cannot move the longitudinal domain wall even in the presence of the magnetic field, and spin orbit torque is needed to explain our experimental data on longitudinal domain wall with current pulses.

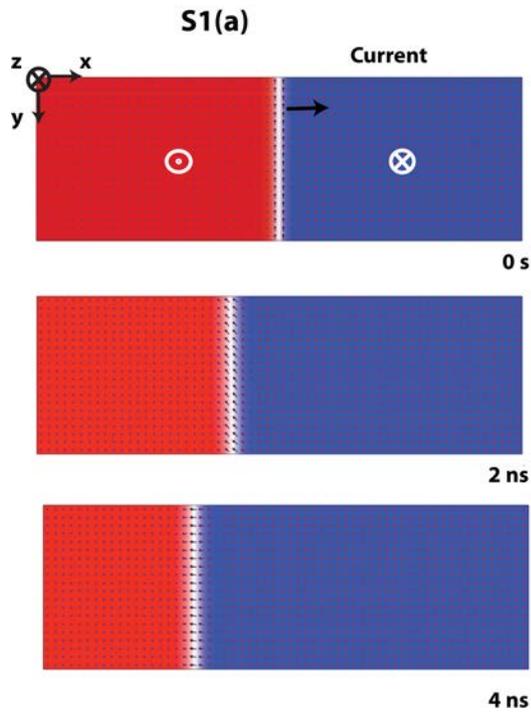

FIG. S1(a) Current flows in the +x direction, i.e. electrons move in the -x direction to apply the bulk spin torque on the transverse domain wall. Blue dots with red background represent moments pointing out of the plane (-z) while red dots with blue background represent moments pointing into the plane (+z). We see that the transverse domain wall moves along the flow of electrons with time as expected.

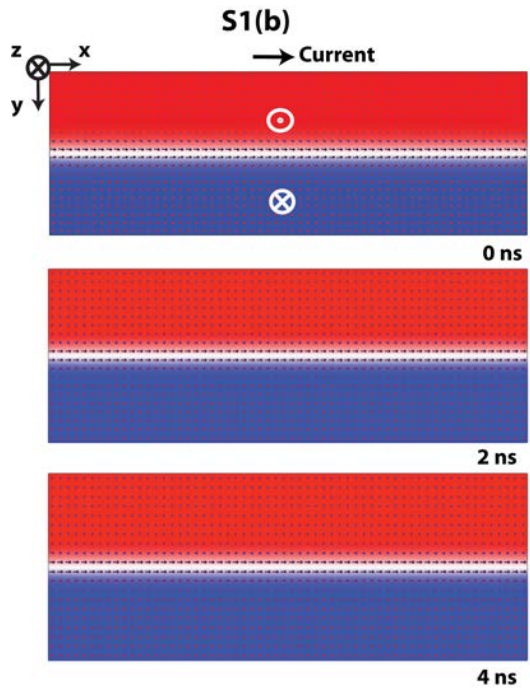

FIG. S1(b) Current flows in +x direction, i.e. electrons move in -x direction, but it cannot apply a bulk spin torque on the longitudinal domain wall. No magnetic field is applied. We see that the longitudinal domain wall does not move with time.

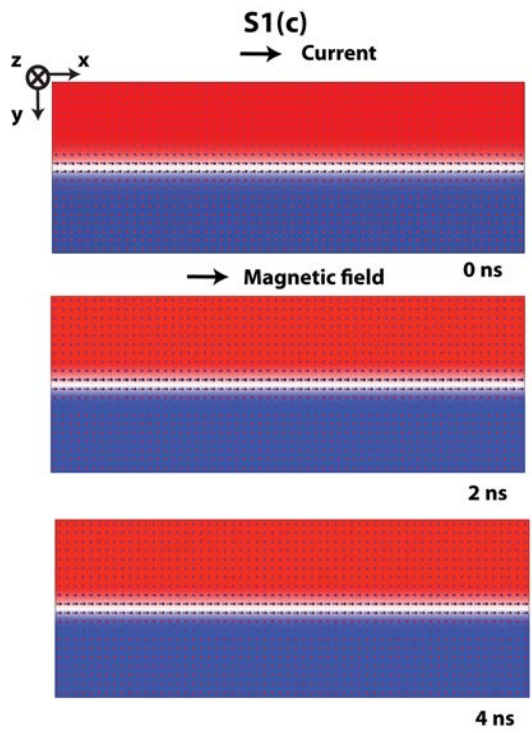

FIG. S1(c) Current flows in +x direction, i.e. electrons move in -x direction. An in-plane magnetic field of 10 G is applied in +x direction. We see that the longitudinal domain wall does not move with time.

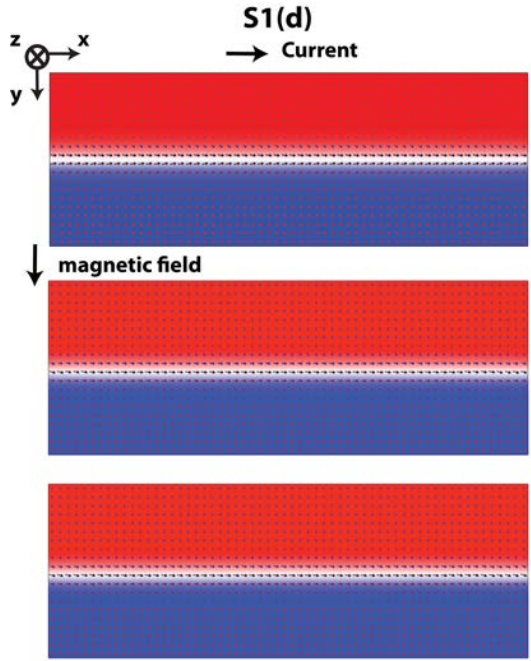

FIG. S1(d) Current flows in +x direction, i.e. electrons move in -x direction to apply the bulk spin torque on the longitudinal domain wall. An in-plane magnetic field of 10 G is applied in +y direction. We see that the longitudinal domain wall does not move with time.

**Section S2- Growth, fabrication and measurement techniques**

Growth and characterization of the materials stack- A thin film stack of Ta (10 nm)/ CoFeB (1 nm)/MgO (1 nm)/ Ta (2 nm) has been sputter deposited on thermally oxidized Si substrate at room temperature. Vibrating Sample Magnetometry is used to characterize the magnetic properties of the thin film stack. When magnetic field is applied out of the plane, the magnet switches sharply close to 0 gauss with the remanent moment close to the saturation moment (Fig. S2). This suggests that out of the plane direction is the easy axis of the magnet. When the magnetic field is applied in plane, a hard axis plot is obtained for magnetic moment versus magnetic field with a field of ~1500 G needed to saturate the magnet in the in-plane direction.

Thus we confirm perpendicular magnetic anisotropy (P.M.A.) in our stack. The saturation magnetic moment is ~20 μ e.m.u measured for a 5 mm by 5 mm thin film sample. Thus the saturation magnetization of the 1 nm thick CoFeB layer is ~ 800 e.m.u./c.c. , or ~$8 \times 10^5$ A/m. This value of saturation magnetization ($M_s$) is used in the micromagnetic simulations.

Fabrication of the device- Orthogonal Hall bars are fabricated from it using optical lithography and ion milling. The bar along x-axis (Fig. 1d) is 500 μm long and 20 μm wide and the bar along y-axis is 500 μm long and 5 μm wide.

Anomalous Hall resistance measurement- Anomalous Hall resistance ($R_{AHE}$) is measured by applying dc current of 100 μA (current density- $5 \times 10^4$ A/cm$^2$) along the bar along the x-axis and measuring the Hall voltage across the bar along y-axis with a nanovoltmeter [Fig. 1(d) of main paper]. The current applied for $R_{AHE}$ measurement is 2 orders of magnitude lower than that used for creating the "mixed" state or moving the domain walls.

Magneto-optic Kerr effect imaging- A Magneto-Optic Kerr Effect (MOKE) microscope is used for magnetic imaging of the Hall bars. The MOKE microscope consists of a 455 nm LED source, two polarizers and a 0.45 NA objective, nominally giving a resolution of 1 μm. To observe contrast, first the magnet is saturated in the "into the plane" or +z direction and an image is taken [Fig. 2(a) of main paper]. This is our reference image. Then a current pulse, a magnetic field or both are applied and another image is taken. Then the two images are aligned and the reference image is subtracted from the other to generate the final MOKE image, which we use for this work. When the bar is saturated "into the plane" the background substrate does not contribute to magnetic signal but the bar does. So when the reference image of an "into the plane" saturated magnet is subtracted from another image of "into the plane" magnet, the final image shows no contrast between the bar and the background [Fig. 2(a)- "into the plane" (⊗) saturated magnet]. When the bar is saturated in the "out of plane" direction, there is no signal from the background substrate but the signal from the bar is strong and negative of the signal from the bar when it is saturated in the "into the plane" direction. As a result when an image of "into the plane" saturated magnet, which is the reference image, is subtracted from the image of "out of the plane" saturated magnet, the resulting MOKE image shows a dark contrast between the bar and the background substrate [Fig. 2(a)-"out of the plane" (⊙) saturated magnet].

Pulsing experiment- Each data point in the plots of Fig. 3b and Fig. 3c of the main paper is obtained by first saturating the magnet to "into the plane" ($m_z=-1$) state, then applying a current pulse of magnitude $7.5\times10^6$ A/cm$^2$ and duration 1 s along the bar in +x direction at zero magnetic field to create the longitudinal domain wall and finally applying another current pulse of a certain magnitude and polarity at a certain in-plane magnetic field. At the end of the pulse anomalous Hall resistance ($R_{AHE}$) is measured and plotted against the current applied with the second pulse and the in-plane magnetic field applied. All the data in this letter are obtained with current pulses of duration 1 s. We have repeated the experiments with current pulses of duration as low as 1 µs to obtain the same results. All the measurements are performed at room temperature.

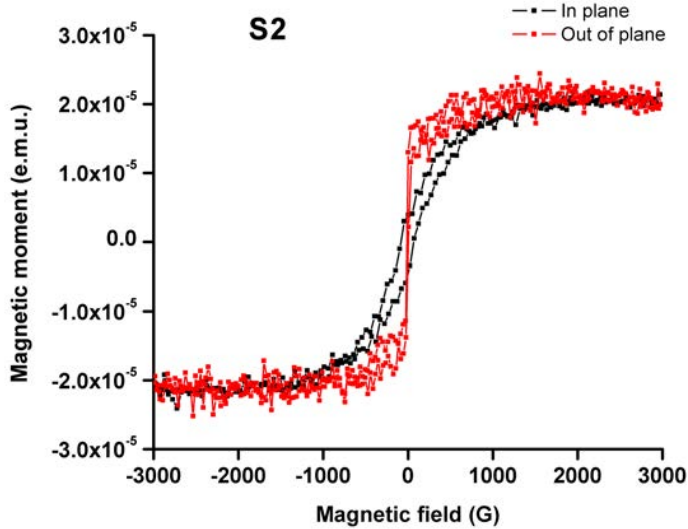

FIG. S2- Magnetic moment versus magnetic field plots for the out of plane direction (red plot) and the in plane direction (black plot) show that the easy axis is in the out of plane direction.

**Section S3- Micromagnetic simulaton showing the formation of the domain wall at the center of the bar due to current pulse**

In Fig. S3(a), using Comsol simulations, we plot the Oersted field generated by the 7.5 x $10^6$ A/cm$^2$ current pulse, which is needed to create the mixed state. We see that the edges experience ~10 G of magnetic field (red and blue patches in Fig. S3(a)) whereas the out of plane component of the magnetic field is negligible anywhere other than the very edges of the bar. The edges also have more defects than the center due to fabrication issues and hence the magnetic material at the edges has lower perpendicular magnetic anisotropy. Hence reverse domains nucleate at the edges of the bar when a current pulse of magnitude $7.5 \times 10^6$ A/cm$^2$ or higher is applied on the bar at a zero magnetic field. Also the Joule heating due to the fairly high magnitude of current pulse can help in the nucleation process.

Next we perform micromagnetic simulations to show how the domain wall moves from the edge to the center of the bar once reverse domains nucleate at the edge of the bar. In Figure S3(b) the magnet is initially saturated "into the plane" ($m_z$=1; red dots in blue background). As current flows in +x direction, Oersted field generated by the current switches the moments near the upper edge of the bar (y<0) to "out of the plane" ($m_z$=-1, blue dots in red background). A longitudinal domain wall is formed as a result near the upper edge of the bar as shown in Fig. S3(b). We simulate a 600 nm long and 200 nm wide magnetic bar using the same simulation parameters as used in Supplementary Section 1 with the initial condition of a longitudinal domain wall near the upper edge of the bar (Fig. S3(b)-image of the magnet at time = 0 ns). We use the standard time evolver provided by OOMMF (Oxs_EulerEvolve) to let the system evolve with time taking into account all the energy terms- exchange energy, anisotropy energy and magnetostatic energy. We observe that the longitudinal domain wall moves in +y direction towards the center with time to reduce the total energy of the system. This is because magnetostatic energy is minimum when the domain wall is at the center of the bar. The final steady state of the system consists of a longitudinal domain wall at the center (y=0) with the upper part of the bar being polarized "out of the plane" and the lower part being polarized "into the plane". The polarity of the domains matches with what we observe in the experiment (Figure 2(c) of the main paper).

Similarly starting from a magnet saturated in the "into the plane" direction current flowing in the –x direction nucleates "out of the plane" polarized domains at the lower edge of the magnet resulting in the formation of a longitudinal domain wall near the lower edge (Supplementary Fig. S3(c)- image of the magnet at time= 0 ns). As time progresses the longitudinal domain wall moves towards the center and at the steady state we get a "mixed state" with equal number of "out of the plane" and "into the plane" polarized domains. The polarity of the domains matches with what we observe in the experiment (Figure 2(c) of the main paper).

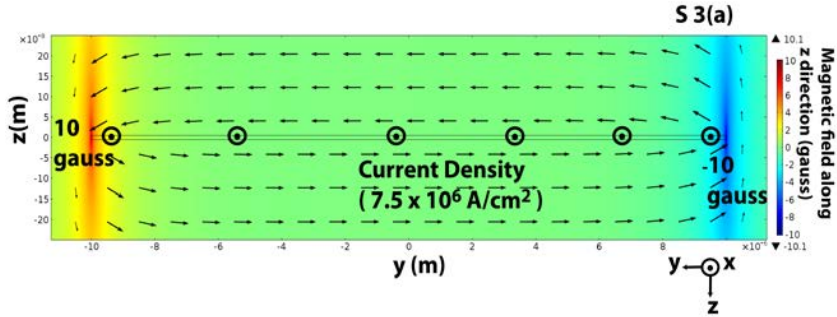

FIG. S3(a) Current flowing along the bar in Figure 2(c) of the main paper is shown to flow out of plane here (⊙). The same coordinate axes of Fig.S3a and S3b are used. x-axis represents the longitudinal direction of the bar, y-axis the transverse direction (width of Ta wire- 20 microns) and z axis the thickness (thickness of Ta wire- 10 nm). We see that the Oersted field is in the plane (y axis) near the center of the bar but it is in the out of the plane direction (+z/-z) near the edge of the bar (red and blue patches). The estimated Oersted field for $7.5 \times 10^6$ A/cm$^2$ of current, used in the simulation, is ~ 10 G near the edge of the bar, as read from the color bar for out of plane component of the magnetic field.

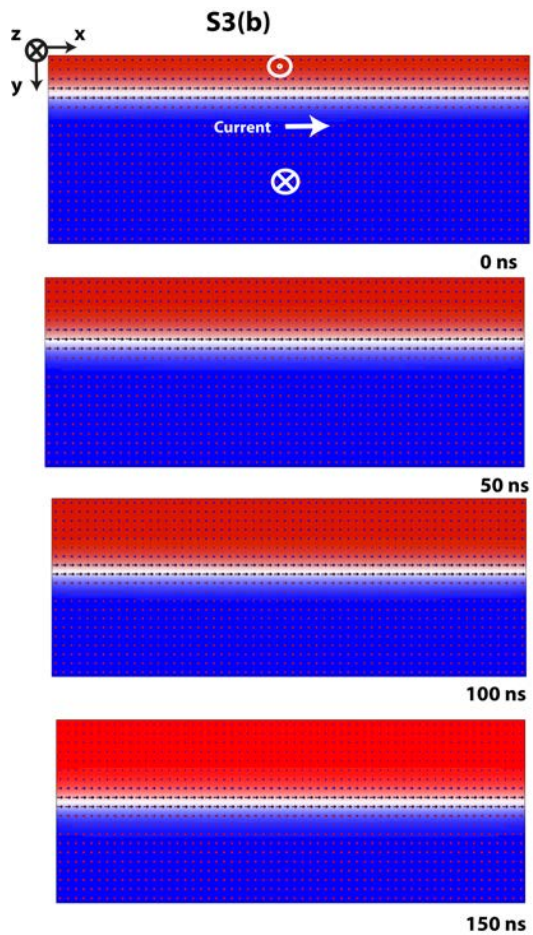

FIG S3(b)- Current in +x direction creates a domain wall at the upper edge of the bar (y<0) because of its Oersted field. Micromagnetic simulation shows that starting from the upper edge of the bar the domain wall moves to the center of the bar at a zero magnetic field to lower its magnetostatic energy.

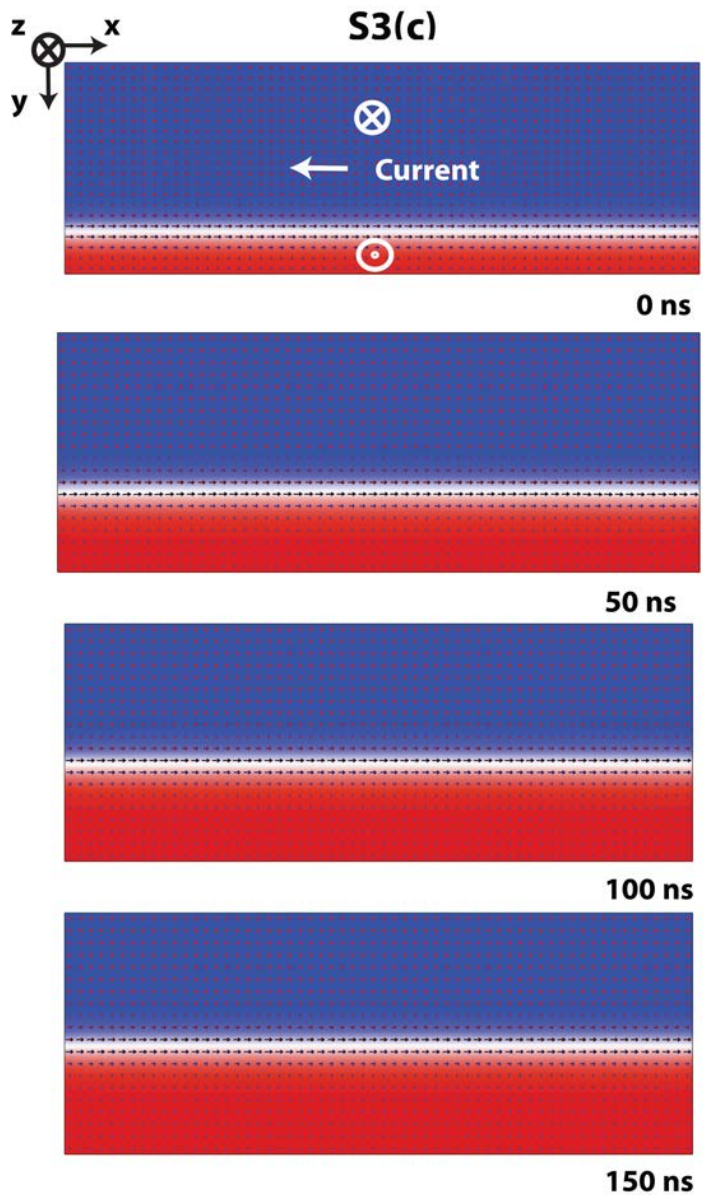

FIG S3(c)- Current in -x direction creates a domain wall at the lower edge of the bar (y>0) because of its Oersted field. Micromagnetic simulation shows that starting from the lower edge of the bar the domain wall moves to the center of the bar at a zero magnetic field to lower its magnetostatic energy.

**Section S4- Details of the method of micromagnetic simulation used to simulate the effect of spin orbit torque on longitudinal domain wall**

Micromagnetic simulations are performed in Object Oriented Micromagnetic Framework (OOMMF) [5] using the spin torque extension module- CYY_STTEvolve [7]. A 600 nm long, 200 nm wide and 1 nm thick magnet is simulated with a 2 nm mesh size laterally and 1 nm mesh size along the thickness. Thus the simulation is basically performed on a two dimensional grid. At every point in the grid (x, y), the moment is allowed to evolve under time following the Landau Lifschitz Gilbert equation with the Slonczewski spin transfer torque term:

$$\frac{d\vec{M}(x,y)}{dt} = -\overline{\gamma}\left(\vec{M}(x,y)\times\overrightarrow{H_{eff}}(x,y)\right) - \frac{\overline{\gamma}\alpha}{M_s}\vec{M}(x,y)\times\left(\vec{M}(x,y)\times\overrightarrow{H_{eff}}(x,y)\right) - \overline{\gamma}\tau(\vec{M}(x,y)\times(\vec{M}(x,y)\times\vec{\sigma})) \quad (1)$$

, where $\vec{M}(x,y)$ is the magnetization at a point in the grid with coordinates (x,y) (Figure 4) and can point in any direction in the (x,y,z) space, $\overrightarrow{H_{eff}}(x,y)$ is the effective field experienced by the magnetization at that point (x,y), α- damping constant, $\overline{\gamma}$ - gyromagnetic ratio, $\vec{\sigma}$- direction of spin polarization and $\tau = \frac{\hbar\theta}{2e\mu_0 M_s t_F}$ [4,8,9], θ- spin orbit torque efficiency, $J_c$- charge current, e- charge of an electron, $\mu_0$- vacuum permeability, $M_s$- saturation magnetization of the ferromagnet ($8\times10^5$ A/m or 800 e.m.u./c.c. used for simulation) and $t_F$- thickness of the ferromagnet ( 1 nm).

The effective field $\vec{H}_{eff}(x,y)$ is calculated using $\overrightarrow{H_{eff}}(x,y) = -\frac{1}{M_s}\nabla_{\vec{M}}E_{total}$, where

the total energy density $E_{total}=E_{anisotropy}+E_{exchange}+E_{Zeeman}+E_{magnetostatic}$

Anisotropy energy $E_{anisotropy}= -K\,M_z^2$, corresponding to the perpendicular magnetic anisotropy of the ferromagnet. K= $6\times10^5$ J m$^{-3}$ = $6\times10^6$ erg/c.c. is used in the simulation.

Exchange energy $E_{exchange}= A((\nabla M_x)^2 + (\nabla M_y)^2 + (\nabla M_z)^2)$, where A is the exchange correlation constant ($3\times10^{-11}$ J m$^{-1}$ or $3\times10^{-6}$ erg cm$^{-1}$ ).

Zeeman energy $E_{Zeeman}= -\mu_0(\vec{M}.\vec{H}_{applied})$ where $\vec{H}_{applied}$ is the applied magnetic field.

$E_{magnetostatic}$ is the magnetostatic energy or dipole energy of the system, calculated by the micromagnetic simulator.

The initial condition of the simulations in Fig. 4 of the main paper is a longitudinal domain wall, which divides the magnet into oppositely polarized domains. To simulate that we use the function $M_z(x,y) = -M_s \cos(\Psi)$; $M_x(x,y) = M_s \sin(\Psi)$ for $H_{applied,x} > 0$ and $M_x(x,y) = -M_s \sin(\Psi)$ for $H_{applied,x} < 0$, where $\Psi = tan^{-1} \sinh\left(\frac{\pi(y-y_0)}{\delta_{DW}}\right)$ [10], $y_0$-position of the domain wall, which is the centre of the bar initially (t=0 s) and $\delta_{DW}$= domain wall width = $\pi\sqrt{\frac{A}{K}}$ = 22 nm.

The system is allowed to evolve using equation (1) under the application of a 10 G magnetic field ($\vec{H}_{applied}$) in +x/-x direction and a spin polarization of $\tau\vec{\sigma}$ where $\tau = \frac{\hbar\theta}{2e\mu_0 M_s t_F}$ and $\vec{\sigma} = -\hat{y}$ (Fig. 4) with θ=0.08, $M_s = 8\times10^5$ A/m = 800 e.m.u./c.c., $t_F$=1 nm and $J_c$=3.5 x $10^6$ A/cm$^2$.

**Section S5- Negligible Dzyaloshinskii-Moriya Interaction (DMI) in our sample**

To measure the magnitude of DMI in our sample we create a transverse magnetic domain wall in our magnetic bars. The $R_{AHE}$ versus out of plane magnetic field plot of Figure 2a (main text) shows that the magnet can be switched by 30 G of magnetic field. However the $R_{AHE}$ measured is only proportional to the average moment in the intersection region of the two Hall bars. MOKE imaging of the switching process shows that a transverse domain wall is nucleated near an end of the bar when an out of plane switching field of magnitude less than 30 G is applied. As the field increases this wall moves along the bar and switches the whole magnet. Such a transverse domain wall is shown in Figure S5(a).

If the DMI of the system is considerable and hence the transverse domain wall is a Neel wall, application of a current pulse along the bar will move it even in the absence of an external magnetic field [4]. So we saturate the magnetic bar in the "into the plane" direction and apply a magnetic field in out of the plane direction, which is less than 30 G. A transverse domain wall nucleates near the right end of the bar every time (Figure S5(a) and S5(b)). Then the magnetic field is turned off. The bar is imaged to record the position of the transverse wall. Then a current pulse is applied along the bar at a zero external magnetic field. The bar is imaged at the end of the current pulse and the position of the domain wall after the current pulse is recorded (Fig. S5). We repeat this experiment several times changing the magnitude and polarity of the current pulse. We observe that the position of the transverse domain wall does not change before and after the current pulse. The domain wall gets distorted after a current pulse of higher magnitude but its average position does not change. The distortion is related to the onset of formation of the "mixed state" because we are applying a current pulse along the bar at a zero magnetic field just like in Figure 2 of the main paper.

From the experimental result that a current pulse across a transverse domain wall does not move it unlike the experiments reported by S. Emori and colleagues [4], we infer that the DMI in our system is negligible.

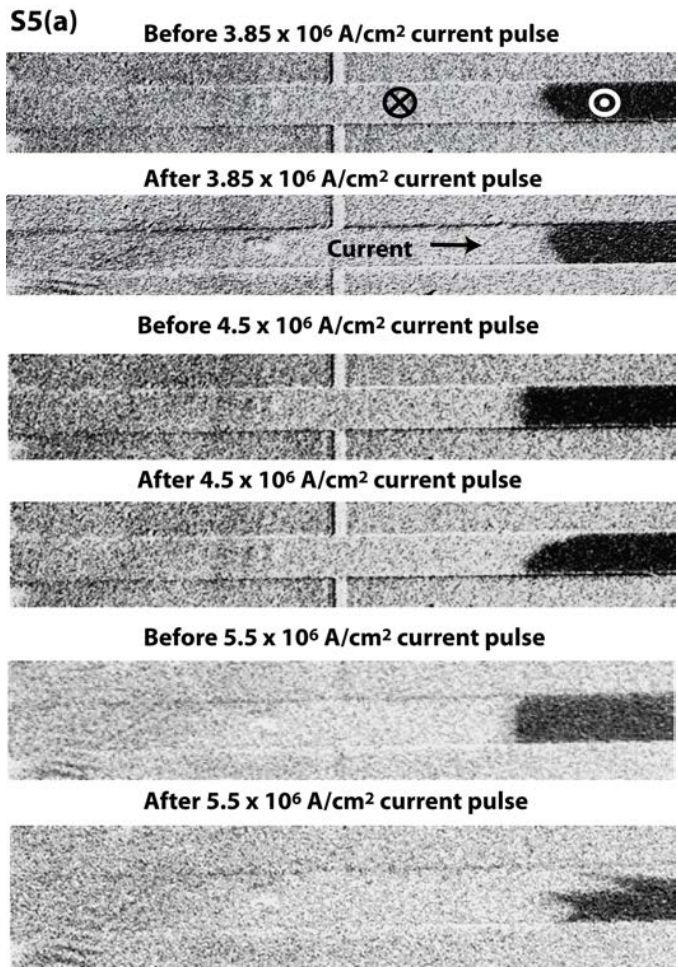

S5(a)
Before 3.85 x 10$^6$ A/cm$^2$ current pulse

After 3.85 x 10$^6$ A/cm$^2$ current pulse
Current →

Before 4.5 x 10$^6$ A/cm$^2$ current pulse

After 4.5 x 10$^6$ A/cm$^2$ current pulse

Before 5.5 x 10$^6$ A/cm$^2$ current pulse

After 5.5 x 10$^6$ A/cm$^2$ current pulse

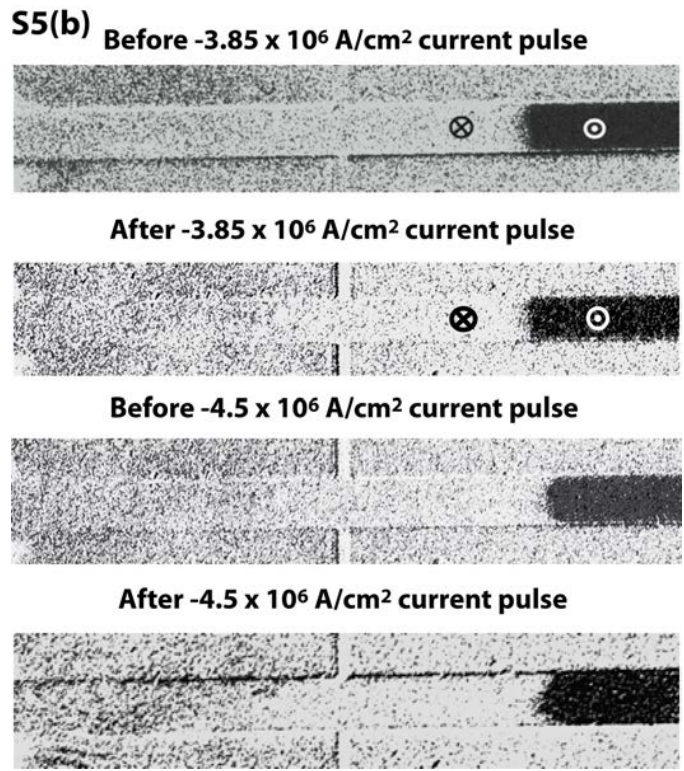

FIG 5a,b MOKE images of the transverse domain wall before and after the application of current pulses of different magnitudes and polarities show no movement of the transverse domain wall due to the current pulses.

**Section S6- Evidence of the domain wall motion not being a result Oersted field of the current or field like spin orbit torque**

We have demonstrated experimentally that a longitudinal domain wall can be moved from the center to the edge of the bar with a current pulse along the bar (Fig. 3). Since the direction of motion of the domain wall is dependent on the polarity of the current, it cannot be attributed to Joule heating.

When a current pulse is applied on the longitudinal domain wall, which is at the center of the bar, the domain wall experiences the Oersted field, generated by the current, which acts in the in-plane direction along the width of the bar (Figure S3c of Supplementary Information). The

Oersted field generated by the current pulse or the field like spin orbit torque from the current cannot explain this kind of domain wall motion as well. The Oersted field or field like torque, if applied along with the external magnetic field, can change the configuration of the moments in the domain wall but cannot move the wall. We performed micromagnetic simulation (Fig. S10) where a field of 10 G is applied in x direction, similar to the external magnetic field in the experiment, and a field of 2.5 G in y direction, which can represent the Oersted field or the field like spin orbit torque from the current pulse. We observe that the domain wall does not move with time under the influence of these two orthogonal fields.

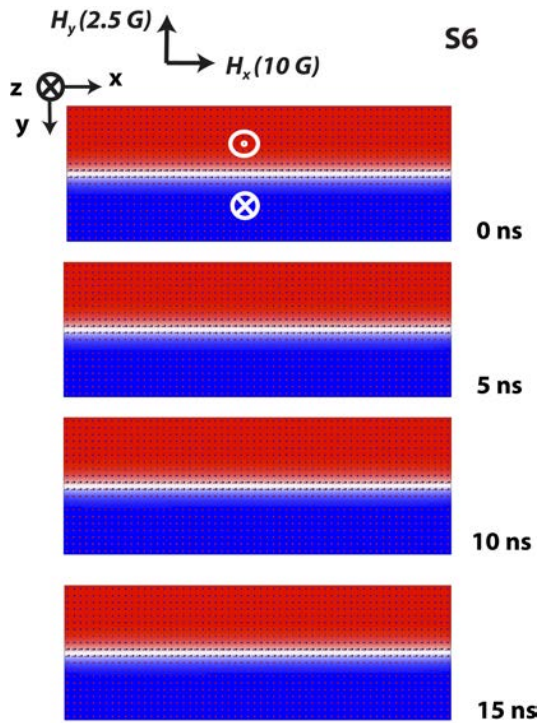

FIG S6- Micromagnetic simulation of time evolution of a longitudinal domain wall under the application of 10 G along x and 2.5 G along y shows no domain wall motion.

**Section S7- Calculation of the spin orbit torque efficiency**

In Fig. 3(c) of the main text we observe that the $R_{AHE}$ increases with the magnitude of the current pulse for a given in-plane magnetic field because a stronger current pulse moves the domain wall further (Fig. 3(a) of main paper). The rate of this increase is low for small magnetic fields but increases with increasing magnitude of field (Fig. S7(a)). For a magnetic field of magnitude 45 G and above (we use negative sign to represent magnetic field in –x direction, so we talk about magnitudes to imply the strength of the field) $R_{AHE}$ increases with the current density of the pulse till it reaches saturation. This corresponds to the longitudinal domain wall moving all the way from the center to the edge of the bar to switch the magnet from "mixed" state to a saturated "into the plane" state (Fig. 3). Using linear fit around the region of the curves where $R_{AHE}$ increases with current, represented by dotted straight lines in Fig. S7(a), we obtain the rate of increase of $R_{AHE}$ with current ($\frac{\partial R_{AHE}}{\partial J_C}$). In Fig. 7(b) we plot $\frac{\partial R_{AHE}}{\partial J_C}$ against the applied in-plane magnetic field. We see that $\frac{\partial R_{AHE}}{\partial J_C}$ increases with the magnitude of applied magnetic field till it reaches saturation for 45 G and above. This happens because a magnetic field of sufficient magnitude (45 G and above in this case) drives the net moment of the domain wall completely in its direction. Hence the "effective out of plane magnetic field" experienced by the domain wall, which is the cross product of net moment in the domain wall and the spin polarization, is the maximum. Stronger the effective field, the higher is the domain wall motion due to the current and higher is the $\frac{\partial R_{AHE}}{\partial J_C}$. We take average of $\frac{\partial R_{AHE}}{\partial J_C}$ values for magnetic fields of magnitude 45 G and above to get a value of

$7.39 \times 10^{-7}$ $\Omega$/(A/cm$^2$).

We have also shown that the longitudinal domain wall can be moved from the center of the bar to the edge with external magnetic field, applied in the out of the plane direction (Fig. 2 of main text). Comparing the change in $R_{AHE}$ corresponding to the domain wall motion due to the externally applied out of plane field with the change in $R_{AHE}$ due to the current pulse would give us an estimate of the "effective out of plane magnetic field" experienced by the domain wall due to the current. Fig. S7(c) shows that $R_{AHE}$ increases with an external magnetic field in the out of the plane direction ($H_{out}$) corresponding to the domain wall motion of Fig. 2b. $R_{AHE}$ finally reaches saturation because the domain wall has moved has moved all the way from the center to the edge. $\frac{\partial R_{AHE}}{\partial H_{out}}$ corresponding to the linear region of the curve is 0.15 $\Omega$/G. Thus, "effective out of plane field" experienced by the domain wall due to the current $\frac{\partial H_{out}}{\partial J_C} = (\frac{\partial R_{AHE}}{\partial J_C})/(\frac{\partial R_{AHE}}{\partial H_{out}})=$ 4.92×10$^{-6}$ G/(A/cm$^2$).

Using the 1D domain wall theory [10], Thiaville and colleagues [11] developed an expression for the "effective out of plane" spin orbit field experienced by a transverse Neel wall when current flows across it. The spin polarization at the interface is orthogonal to the net magnetic moment of the Neel wall and applies an effective out of plane field $H_{out} = \frac{\pi}{2}\frac{\hbar\theta}{2et_F M_s}$, on the domain wall, where θ is the spin orbit torque efficiency, e is the charge of an electron, $\hbar$ is the Planck constant, $\mu_0$ is the vacuum permeability, $M_s$ is the saturated magnetization of the ferromagnet ($8\times10^5$ A/m, as measured for our thin film stack by Vibrating Sample Magnetometry) and $t_F$ is the thickness of the ferromagnet (1 nm in our case). The longitudinal domain wall in our experiment is a Bloch wall and the net magnetic moment of the wall is along the direction of the applied in-plane field. Since current flows along the longitudinal wall, the spin polarization ($\vec{\sigma}$) at the interface due to SHE is orthogonal to the net magnetic moment of the domain wall (black arrow of Fig. 4) as shown in Fig. 4a and Fig. 4b of the main paper. So the same expression for the "effective out of plane field" can be used for our experiment. The experimental value of $\frac{\partial H_{out}}{\partial J_C}$ is used in the expression to extract the spin Hall angle ($\theta_{SHE}$), which turns out to be equal to 0.076.

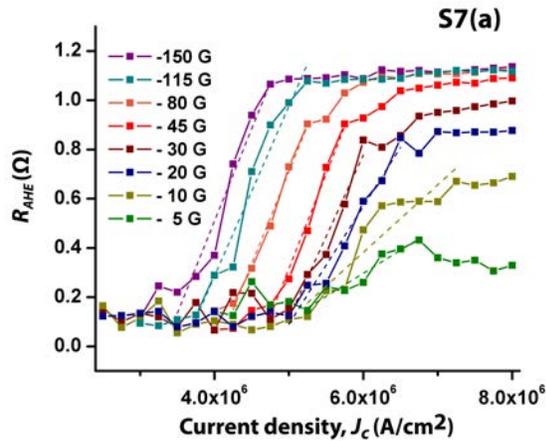

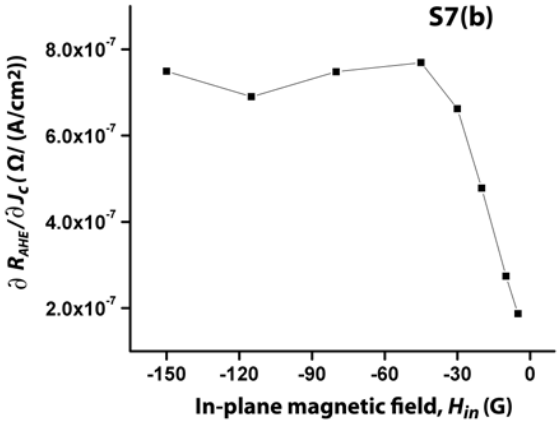

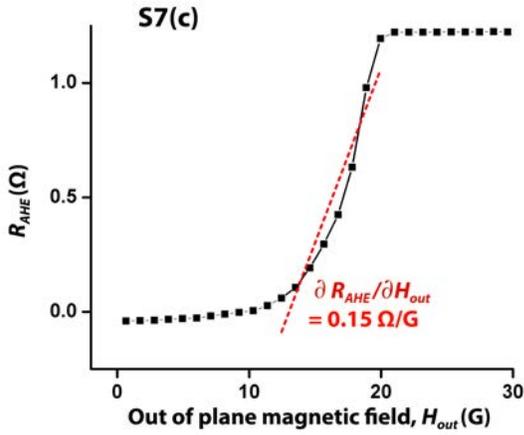

FIG 7(a) Starting from a longitudinal domain wall at the center of the bar every time, current pulses of different magnitude are applied for an applied in-plane magnetic field in –x direction and $R_{AHE}$ is measured after every pulse. $R_{AHE}$ is plotted against the current density of the current pulses for different in-plane fields. Using linear fit for the region where $R_{AHE}$ increases with the current, $\frac{\partial R_{AHE}}{\partial J_C}$ values (slopes of the dotted lines) for different in-plane fields are obtained. (b) Plot of $\frac{\partial R_{AHE}}{\partial J_C}$, obtained in Fig. 10a versus applied in-plane field. (c) Starting from a longitudinal wall at the center of the bar, out of plane magnetic field ($H_{out}$) is applied of increasing magnitude. The $R_{AHE}$ increases with increasing field corresponding to the domain wall moving from the center of the bar to edge to switch the magnet from "mixed" state to saturated "into the plane"

state (Fig. 2a). Using a linear fit for the region where $R_{AHE}$ increases with $H_{out}$, $\frac{\partial R_{AHE}}{\partial H_{out}}$ (slope of the dotted line) is obtained.